\pdfoutput=1
\documentclass[11pt,a4paper]{article}
\usepackage[T1]{fontenc}
\usepackage[utf8]{inputenc}
\usepackage{lmodern}
\usepackage[margin=2.5cm]{geometry}
\usepackage{booktabs,tabularx,array}
\usepackage{microtype}
\usepackage[hidelinks]{hyperref}
\usepackage{url}
\usepackage{tikz}
\usetikzlibrary{positioning,arrows.meta,decorations.pathreplacing}
\usepackage{pgfplots}
\pgfplotsset{compat=1.18}
\definecolor{err}{HTML}{EB6834}
\definecolor{cor}{HTML}{2A78D6}
\definecolor{nei}{HTML}{C9C8C2}
\definecolor{ink2}{HTML}{52514E}

\newcolumntype{Y}{>{\raggedright\arraybackslash}X}
\newcolumntype{P}[1]{>{\raggedright\arraybackslash}p{#1}}
\title{Officially corrected errors persist in AI training data\\[0.3em]
\large Preregistered measurements on EU law, Internet standards and retracted science}
\author{Sébastien Gracia\\
\small Loxyn, Lyon, France\\
\small ORCID \href{https://orcid.org/0009-0001-7885-2655}{0009-0001-7885-2655} \quad \texttt{sebastien@loxyn.ai}}
\date{October 2026}

\begin{document}
\maketitle

\begin{abstract}
Official texts are corrected after publication: EU corrigenda, verified RFC errata, retractions.
We measured whether these corrections are present in four public datasets used to train language
models, and what models answer about the corrected passages. In EurlexResources, a legal training
dataset, 1,989 of 2,082 testable single-language corrections to EU acts (95.5\%, 94.6--96.3) remain in
their erroneous form; in a random sample of 100, 46 change the meaning, a figure, a cross-reference,
the competent authority or the scope (20 without scope). At least 124 corrected EU-law errors and at
least 60 verified RFC errata (38 classified ``Technical'') are reproduced in their erroneous form in
FineWeb-2 and in a sample of FineWeb. About 12,200 retracted papers (11,700--12,500) are present in
the peS2o corpus, 79\% with no mention of the retraction; separately, 53\% of a random sample of
retracted papers are still cited at least a year after retraction (OpenAlex). Asked 59
questions on serious corrected errors without documents, five models gave the erroneous version in
45\% of the answers in which they committed to one version (31--60\% allowing for clustering by
question): 53\% for scope corrections, 32\% for the others. One answer in 295 said it did not know.
Shown both versions side by side, the models chose the corrected one in 53\%
of cases (49--57), and naming the act made no difference; when each version carried its date and the
corrigendum label, the four larger models chose it in 99--100\% of cases. A
comparison between a legal masked language model and its base model is null, against our bet: we do
not show that model answers come from the datasets measured. All steps were preregistered and their
fingerprints timestamped in the Bitcoin blockchain; six controls failed, several bets were lost, and
every deviation is reported. Data and proofs: \href{https://doi.org/10.5281/zenodo.23642239}{doi:10.5281/zenodo.23642239}.
\end{abstract}

\section{Introduction}

An officially corrected error should stop circulating. Three institutions publish such corrections in
a structured way. The European Union publishes corrigenda to its acts in the Official Journal; a
corrigendum does not change the original publication but states what should be read instead, and the
consolidated versions on EUR-Lex include it. The RFC Editor publishes errata to Internet standards and
marks some as ``Verified''. Scientific journals retract papers, and the Retraction Watch database,
now distributed by Crossref, lists the retractions.

Language models are trained on text collected from the web and from specialised sources. When those
sources hold the text as first published, the correction is absent unless the collector added it. We
asked three questions:
\begin{enumerate}
\item Are corrected errors present, in their erroneous form, in public training datasets?
\item Are the corrections serious?
\item What do models answer about the corrected passages, and can they tell which version applies?
\end{enumerate}

\textbf{Related work.} Corrigenda to EU law have been studied by legal linguists
\cite{bobek2009,prietoramos2020,bielpytel2021}. Corrections to an evaluation dataset fail to reach the
data that models are trained on \cite{conllsharp2024}. Training corpora have been inspected for
content and provenance \cite{elazar2024,longpre2023}, and documentation practices for datasets have
been proposed \cite{gebru2021}; we add a check against official corrections. Language models learn
long-tail facts poorly \cite{kandpal2023} and their knowledge ages \cite{lazaridou2021,dhingra2022};
legal AI tools make measured errors \cite{dahl2024,magesh2024}. Retracted papers keep being cited
\cite{barilan2017,hsiao2021}, and chat models do not flag retractions \cite{thelwall2025,llmretracted2026}.
To our knowledge, the presence of corrections to EU law and to Internet standards in training
datasets, and the count of retracted papers in a named training dataset, had not been measured.

\section{Method}

\begin{figure}[t]\centering\small
\begin{tikzpicture}[
  box/.style={draw=ink2,rounded corners=2pt,align=center,inner sep=3pt,minimum height=1.3cm,text width=2.45cm,font=\footnotesize,execute at begin node={\hyphenpenalty=10000}},
  arr/.style={-{Stealth[length=5pt]},thick,ink2}]
\node[box] (a) {\textbf{Official correction}\\EU corrigendum,\\RFC erratum,\\retraction};
\node[box,right=0.3cm of a] (b) {\textbf{Probe}\\erroneous text\\vs corrected text};
\node[box,right=0.3cm of b] (c) {\textbf{Search}\\in 4 training\\datasets};
\node[box,right=0.3cm of c] (d) {\textbf{Reading}\\each hit checked\\by a reader};
\node[box,right=0.3cm of d] (e) {\textbf{Control}\\random sample,\\bar fixed\\in advance};
\draw[arr] (a)--(b); \draw[arr] (b)--(c); \draw[arr] (c)--(d); \draw[arr] (d)--(e);
\node[box,below=0.7cm of c,text width=10.5cm,minimum height=0.9cm] (m) {\textbf{Model test}: 59 questions on serious corrections, asked\\without documents to 5 models; answers judged blind};
\draw[arr] (b.south) -- (b.south |- m.north);
\draw[decorate,decoration={brace,amplitude=5pt},ink2] ([yshift=0.25cm]a.north west) -- ([yshift=0.25cm]e.north east)
  node[midway,above=6pt,font=\footnotesize,align=center]{written, hashed and timestamped in a Bitcoin block \emph{before} the run; every deviation logged};
\end{tikzpicture}
\caption{Method common to all steps.}
\label{fig:method}
\end{figure}
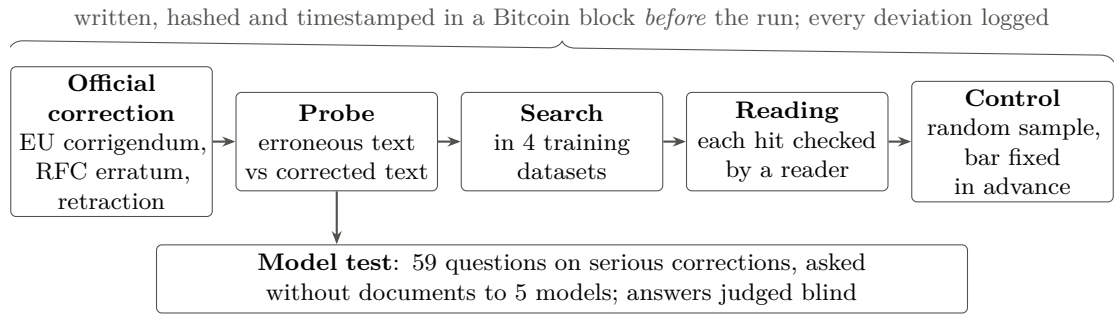

Figure~\ref{fig:method} summarises the method.

\subsection{Sources}

\textbf{Corrections.} From the EU corrigenda that correct a single language version (French, Italian
or English), we extracted 5,673 corrections, of which 5,445 are replacements with a before and an
after text. The extractor was validated before use (recall 27/30 on a fixed test set; coverage 88.7\%
of readable texts, 84.7\% if 125 texts excluded after inspection are counted). The legal-dataset step
uses the 3,847 corrections to acts published up to the end of 2021. For standards, we used the RFC
Editor errata with status ``Verified''. For science, we used the Retraction Watch database (version of
24 October 2026). None of the corrections are ours.

\textbf{Datasets}, at fixed public revisions on Hugging Face: EurlexResources \cite{niklaus2023}
(\texttt{4d7946e}), part of MultiLegalPile; FineWeb-2 \cite{fineweb2} (\texttt{af9c133}; Italian and
French, 599 million pages); a sample of FineWeb \cite{fineweb} (\texttt{sample/100BT},
\texttt{9bb295d}, 147.6 million pages); and peS2o v2 \cite{pes2o} (\texttt{636a503}), the scientific
component of the Dolma corpus \cite{dolma2024}. Citation links came from the OpenAlex API
\cite{openalex}, queried on 24 October 2026.

\textbf{Models.} For the answer test: Mistral-7B-Instruct-v0.3 (local, 8-bit) and, through Scaleway
Generative APIs on 24 October 2026, Mistral Small 3.2 (24B), Mistral Medium 3.5 (128B), Llama 3.3
70B Instruct and Qwen3 235B A22B Instruct. The same five were used on 25 October 2026 for the
side-by-side test, Mistral 7B then in 8-bit GGUF through Ollama. For the likelihood test: legal-xlm-roberta-large
\cite{niklaus2023} and its base model, xlm-roberta-large.

\subsection{Preregistration and timestamps}

Each step was written down before it was run: question, data, measures, control bar and bets. The
text and the scripts were hashed (SHA-256), and the hash file was timestamped with OpenTimestamps,
which anchors it in a Bitcoin block. The anchor proves that the file existed at the latest at the time
of the block. For the legal-dataset, standards, retraction and answer steps, the block precedes the
first results. For three steps (severity, web census, legal model), the block came 5 to 22 minutes
after the first results; their precedence rests only on our logs. Each deposited step includes its
log. Table~\ref{tab:deviations} lists every deviation from the preregistrations.

\subsection{Readers and controls}

Readings, classifications, question writing and judgments were made by a single AI model (Claude, by
Anthropic) in separate sessions that did not see earlier readings, and were checked by further
readings of the same model against bars fixed in advance (typically 27 of 30 cases drawn at random).
The readers had the official texts in front of them; the models tested did not. The review of the
severity classes by a lawyer, planned in the preregistration, was replaced by a review by two other
AI systems (ChatGPT and Gemini) that saw our classifications. A law student, who wished to remain
anonymous, reviewed the legal examples (Table~\ref{tab:law}), six model answers and three sentences of the
discussion, and confirmed all of them, with one nuance taken up in Section~\ref{sec:discussion}. The other
cases were not reviewed by a human. This is the main limit of the study, and the full readings are published so that
they can be checked.

\section{Results}

Table~\ref{tab:results} summarises the measurements.

\begin{table}[h]
\centering\small
\begin{tabularx}{\textwidth}{@{}P{2.2cm}YP{3.3cm}P{2.1cm}@{}}
\toprule
Domain & What we counted & Result & 95\% interval \\
\midrule
EU law & errors still uncorrected in the legal dataset & 1,989 / 2,082 (95.5\%) & 94.6--96.3 \\
 & serious corrections (random 100) & 46 (20 without scope) & 37--56 \\
 & corrected errors copied on the web & $\ge$124 / 1,764 (7.0\%) & 5.9--8.3 \\
\addlinespace
Standards & verified errata copied on the web & 60 / 2,321 (2.6\%) & 2.0--3.3 \\
\addlinespace
Science & retracted papers in peS2o & $\approx$12,229 & 11,694--12,470 \\
 & \quad of which not flagged & 79\% & 72--85 \\
 & still cited $\ge$1 year after retraction & 212 / 397 (53\%) & 48--58 \\
\addlinespace
Models & erroneous version, decisive answers & 57 / 127 (45\%) & 31--60 \\
 & corrected version chosen, side by side & 53.4\% & 49.5--57.2 \\
 & \quad same, labelled with dates & 92.9\% & 91.4--94.4 \\
 & legal model prefers the error more & no ($-1.6$ points) & $p=0.21$ \\
\bottomrule
\end{tabularx}
\caption{Main measurements. Wilson intervals, except for model answers (bootstrap over questions).}
\label{tab:results}
\end{table}

\subsection{EU law in a legal training dataset}

EurlexResources holds EU acts as published. We found no corrigendum in it (no corrigendum CELEX
number among 337,204 documents in three files checked). We searched each corrected passage in the act
it corrects. Among the 2,960 corrections whose act is served, the search can tell the erroneous form
from the corrected one for 2,082 (70.3\%; only 199 of 457 in English). Of these 2,082, 1,989 are
present in their erroneous form (95.5\%, 94.6--96.3): 95.4\% in French (839/879), 95.0\% in Italian
(954/1,004), 98.5\% in English (196/199). The first control of the search rule failed (24/30); the
rule was lengthened after reading, without a new seal, then checked on a new sample (30/30). The
95.5\% figure depends on this modified rule.

\subsection{Severity}

We drew 100 of these errors at random and classified each correction into eight classes, with two
readers and an arbitration (agreement 95.8\% on serious versus not serious, $\kappa=0.92$; both
readers being the same model). Forty-six (37--56) change the meaning, a figure, a cross-reference, the
competent authority or the scope of the provision. Scope is the most debatable class; without it, 20
remain. Of the 46, 21 come from acts in force on 25 October 2026 according to the Publications
Office SPARQL endpoint (property \texttt{resource\_legal\_in-force}; script deposited).
Table~\ref{tab:law} gives examples, all from acts in force.

\begin{table}[h]
\centering\small
\begin{tabularx}{\textwidth}{@{}P{4.3cm}YY@{}}
\toprule
Act (language) & Published text & Corrected text \\
\midrule
Directive 2002/15/EC, Art.~9(b) (FR) & records kept ``at least one year'' & ``at least two years'' \\
Directive 2014/95/EU, Art.~19a (IT) & average headcount equal to 500 & exceeding an average of 500 \\
Regulation (EC) 767/2008, Art.~34 (IT) & records kept by the Commission & by the Management Authority \\
Directive 2011/65/EU, Art.~2(4)(e) (IT) & large-scale industrial fixed installations & large-scale fixed installations \\
Directive 96/9/EC, Art.~7 (IT) & obtaining, verifying and presenting & obtaining, verifying or presenting \\
\bottomrule
\end{tabularx}
\caption{Corrected errors in acts in force, present in their erroneous form in EurlexResources and on
the web. Italian passages are translated.}
\label{tab:law}
\end{table}

\subsection{The web}

A first attempt to measure a circulation rate in FineWeb-2 failed its control twice and is not
published as a measurement. We replaced it with a census: for each correction with candidate pages,
up to five pages were read in random order until one was confirmed to reproduce the erroneous passage
(209 pages read). At least 124 of 1,764 corrections searched (7.0\%) are reproduced; after a further
control the count is about 116. The rate is 11.0\% for Italian (104/943, 69 sites) and 2.4\% for French
(20/821, 18 sites). The 124 corrections come from 78 acts. The confirmed pages span 85 sites, and 113
of the 124 (91\%) are outside EU websites.

\subsection{Internet standards}

Each of the 2,321 verified errata with a usable probe was searched in the FineWeb sample. For 60 of
them (2.6\%), on 48 RFCs, the erroneous passage is still copied, on 20 sites, in pages crawled more than
a month after the erratum was reported. Thirty-eight of these errata are classified ``Technical'' by the
IETF (38 of 1,247 Technical errata). For only 2 of the 60 does the corrected text also appear in a page
crawled after the report. The date used is the report date; a stricter variant, preregistered, using
the last update date of the erratum keeps only 2 errata (0 Technical), mostly because many errata were
updated in a 2019 migration. Examples: RFC 3390 (erratum 4583), where a TCP may start with a
``larger'' instead of a ``smaller'' initial window; RFC 1534 (erratum 545), where the DHCP message type
option is given as 51 instead of 53.

\subsection{Retracted science}

We matched the titles and years of 38,422 retracted papers (retracted before December 2022, with a
unique title of at least 30 characters) against peS2o. The first control of the matching failed
(25/30). Without changing the rule, we measured its precision on 150 new cases (145/150; 92.4--98.6),
which gives about 12,229 retracted papers in the corpus (11,694--12,470), about one in three of those
searched. In health and biology the share is higher: about 8,831 of 18,553 (48\%). In 79\% of them
(115/145; 72--85) nothing in the corpus mentions the retraction. Retraction Watch notices often give
several reasons; among the 12,651 raw matches, the notice cites errors or unreliable results for 27\%,
a paper mill or manipulated peer review for 22\%, duplicated or manipulated images for 17\%,
plagiarism for 11\%, duplicate publication for 10\%, and falsification or fabrication of data for 7\%
(a descriptive calculation made afterwards; script deposited).

An attempt to count, inside peS2o, papers that cite a retracted paper after its retraction failed:
the corpus keeps almost no reference lists (3 of 3,001 papers checked have a references section),
which we had not checked beforehand, and the control failed (19/30). We repeated the measurement on
OpenAlex with a new preregistration. Of 400 retracted papers drawn at random from 52,975 retracted at
least two years earlier, OpenAlex finds 397; 212 of them (53\%, 48--58) are cited by at least one work
published a year or more after the retraction. These papers receive 2,210 such citations (mean 5.6,
bootstrap interval 3.2--9.4, median 1), of which 637 go to a single paper. OpenAlex itself flags 392
of the 397 as retracted. A reading of how these citations use the retracted paper failed its control
(24 of 32 citations confirmed against a bar of 28, mainly because publisher sites blocked the reader).
In the 24 confirmed citations, none mentions the retraction and 15 rely on the paper's results; these
are reported as cases, not as a measurement.

\subsection{Model answers}

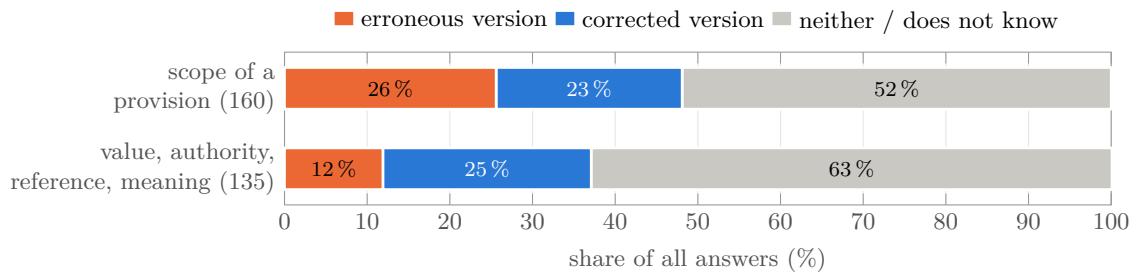
\begin{figure}[t]\centering\small
\begin{tikzpicture}
\begin{axis}[xbar stacked, width=12.5cm, height=3.6cm, bar width=0.55cm,
  xmin=0, xmax=100, xlabel={share of all answers (\%)}, xlabel style={font=\footnotesize,text=ink2},
  symbolic y coords={Scope,Other}, ytick={Scope,Other}, y dir=reverse, clip=false,
  yticklabels={{scope of a\\provision (160)},{value, authority,\\reference, meaning (135)}},
  yticklabel style={align=right,font=\footnotesize},
  axis line style={ink2!60}, tick style={ink2!60}, ticklabel style={font=\footnotesize,text=ink2},
  xmajorgrids, grid style={ink2!15}, enlarge y limits=0.45,
  legend style={at={(0.5,1.06)},anchor=south,legend columns=3,draw=none,font=\footnotesize},
  legend cell align=left,
  nodes near coords={\pgfmathprintnumber[fixed,precision=0]{\pgfplotspointmeta}\,\%},
  every node near coord/.append style={font=\scriptsize,text=black},
  point meta=explicit]
\addplot[fill=err,draw=white,line width=1pt] coordinates {(25.6,Scope) [25.6] (11.9,Other) [11.9]};
\addplot[fill=cor,draw=white,line width=1pt,nodes near coords style={text=white}] coordinates {(22.5,Scope) [22.5] (25.2,Other) [25.2]};
\addplot[fill=nei,draw=white,line width=1pt] coordinates {(51.9,Scope) [51.9] (63.0,Other) [63.0]};
\legend{erroneous version,corrected version,neither / does not know}
\end{axis}
\end{tikzpicture}
\caption{Answers of five models to 59 questions, by kind of correction. Among the answers that commit to one version, the erroneous one is given in 53\% of cases for scope corrections and in 32\% for the others.}
\label{fig:answers}
\end{figure}

From the serious corrections, we wrote 59 questions whose answer differs between the published and the
corrected text, validated each against both texts, and asked them without documents to five models.
A judge that did not know which answer came from which text classified each response (agreement with a
second judgment: 58/60). In 167 of 295 answers (57\%), the models gave neither version. When they
committed to one, they gave the erroneous version in 57 of 127 answers (45\%; 36--54 treating answers
as independent, 31--60 with a bootstrap over questions). The rate depends on the kind of correction:
41 of 77 decisive answers (53\%) on scope corrections, 16 of 50 (32\%) on values, authorities,
cross-references and meaning. One answer in 295 said explicitly that the model did not know. The rate
does not differ visibly by model size (Table~\ref{tab:models}; Figure~\ref{fig:answers}). On three questions all five models
gave the erroneous version, and on six all five gave the corrected one. For instance, on Annex~III,
point~4 of Regulation (EU) No 10/2011 on plastic materials in contact with food, all answered that the
test in simulant D2 is not mandatory, as in the published text, whereas the corrected text makes it
mandatory.

\begin{table}[h]
\centering\small
\begin{tabular}{@{}lccc@{}}
\toprule
Model & Erroneous & Corrected & Erroneous among decisive \\
\midrule
Mistral 7B Instruct v0.3 (local, 8-bit) & 9 & 11 & 45\% (26--66\%) \\
Mistral Small 3.2 (24B) & 10 & 13 & 43\% (26--63\%) \\
Llama 3.3 70B Instruct & 10 & 17 & 37\% (22--56\%) \\
Mistral Medium 3.5 (128B) & 14 & 15 & 48\% (31--66\%) \\
Qwen3 235B A22B Instruct & 14 & 14 & 50\% (33--67\%) \\
\midrule
All & 57 & 70 & 45\% (36--54\%) \\
\bottomrule
\end{tabular}
\caption{Answers to 59 questions without documents. The remaining 168 answers gave neither version;
one of them said it did not know.}
\label{tab:models}
\end{table}

\subsection{Both versions side by side}

At the answer test, the erroneous share among decisive answers (45\%) is close to what a choice at
random between the two versions would give. To see whether the models can tell which version applies,
we showed the same five models, for the same 59 corrections, the published and the corrected passage
and asked which one is correct and applies, answering A or B, in three conditions: the two passages
alone; with the act's reference and the location of the passage; and, in addition, each passage
labelled ``original text, act of \emph{date}'' or ``corrigendum of \emph{date}''. Each pair was asked
in both orders (1,770 answers; readable: 98.8, 99.7 and 98.5\%, against a bar of 95\%).

Unlabelled, the corrected version is chosen in 53.4\% of cases (49.5--57.2, bootstrap over
questions); with the act named, in 53.4\% as well (difference 0.0 points, $-1.6$ to $+1.4$). The
choice then follows position rather than content: the four larger models answer ``B'' in 82--95\% of
cases, and choose the same version in both orders for only 44 of 236 pairs. Labelled, the corrected
version is chosen in 92.9\% of cases (91.4--94.4): 99--100\% for the four larger models, with the same
version in both orders for 235 of 236 pairs, and 65\% for Mistral 7B. Where a model had given the
erroneous version at the answer test, it chooses the corrected one in 48\% of cases unlabelled and
91\% labelled.

The models thus lack the information of which version applies, and use it when it is given. The test
does not tell whether they saw both versions without a way to separate them or neither; nor does it
show what a label in training data would do, since the label here is in the prompt.

\subsection{A null result}

We compared, on 1,602 pairs of published and corrected passages, the pseudo-log-likelihood
\cite{salazar2020} assigned by a legal masked language model trained on MultiLegalPile and by its base
model. Our bet was that the legal model would prefer the erroneous version at least 5 points more
often. It does not: it prefers it in 53.0\% of pairs, the base model in 54.6\% (McNemar $p=0.21$). With
the secondary per-token score, the legal model prefers the erroneous version less often ($-4.1$ points,
$p=0.001$). On the 1,334 cleanest pairs, both models prefer the corrected version slightly more often
than not (the erroneous one in 44.2\% and 46.0\% of pairs). We therefore do not show that the models'
answers come from the datasets measured.

\section{Deviations and bets}

\begin{table}[h]
\centering\small
\begin{tabularx}{\textwidth}{@{}P{3.2cm}Y@{}}
\toprule
Step & Deviation \\
\midrule
Extraction & recall 27/30, bet of 28 lost; coverage 84.7\% if 125 excluded texts are counted \\
Legal dataset & control failed (24/30); rule changed after reading, not sealed; 30/30 after \\
Web, rate & control failed twice; replaced by a census; bet ($\ge$20\%) lost \\
Web, severity, legal model & Bitcoin block 5--22 minutes after the first results \\
Severity & lawyer review replaced by ChatGPT and Gemini, not blind \\
Legal model & bet ($\ge$+5 points) lost; code optimised and part of the run moved to another machine (952/952 identical) \\
Model answers & two one-line code fixes \\
Side by side & bet ($\ge$55\% unlabelled) lost; Mistral 7B quantised as GGUF, not bitsandbytes; answer reader missed ``A (\ldots)'' (18 of 1,770 answers; labelled share 92.9\% becomes 93.7\%) \\
Standards & strict date variant: 2 errata, none Technical \\
Retractions & control failed (25/30); precision measured on 150 new cases \\
Cascade, peS2o & no reference lists in the corpus; control failed (19/30) \\
Cascade, OpenAlex & 32 citations read instead of 40; control failed (24/32) \\
\bottomrule
\end{tabularx}
\caption{Deviations from the preregistrations, each dated in the step's log.}
\label{tab:deviations}
\end{table}

Some bets were also set when part of the result was known (for instance, that the legal dataset holds
no corrigenda). Preregistration here establishes traceability, not bold predictions.

\section{Discussion}
\label{sec:discussion}

The erroneous version is not an invention: it is the original official text, which stays published
and accessible. It does not take precedence, since the corrected text is the one that applies, but a
reader who checks an answer against the original text finds the same error. Only the corrigendum, the
consolidated version, the erratum or the retraction notice reveals it, and these are what the
datasets measured lack. The consequences were not measured. Plausible ones are a professional
retaining a wrong period, threshold or authority from an assistant; a retracted meta-analysis reused
in a review; software implementing the wrong value of a standard; and generated texts that repeat the
erroneous version and are collected in turn.

The remedies are metadata. A document-level field pointing to the corrigendum or the consolidated
version (EUR-Lex already links them), to the verified erratum (the RFC Editor publishes errata in
machine-readable form), or to the retraction notice (Crossref already exposes retractions) would let
dataset builders drop, replace or flag corrected texts. In a prompt, such a label is enough for the
larger models to choose the corrected version; we did not measure its effect in training data. Model
developers can measure answers on
corrected texts and train models to say they do not know.

\section{Limits}

\begin{itemize}
\item All readings and judgments were made by one AI model; only the examples were reviewed by a
person, a law student.
\item Law: three languages for the legal dataset, two for the web, and only corrections made in a
single language; the search can test 70\% of the corrections.
\item Standards: a sample of about 1/150 of English FineWeb; the result depends on the date used.
\item All counts of copies are minimums.
\item The answer test has 59 questions and no control questions on never-corrected passages; most
erroneous answers concern scope corrections, the most debatable class; models were run through a
host, not their vendor, and the short prompt pushes them to answer. These are the conditions of a user
asking without documents, not of products that rely on retrieval.
\item In the side-by-side test, the labelled condition is easy: the prompt says that one version
comes from a corrigendum, and the label says which.
\item Retractions are matched by title and year; about 3\% of matches point to another publication of
the same work. OpenAlex citation links are imperfect (2 of 32 wrong in our reading).
\item We measure what is in the data and what models answer, not decisions taken on those answers.
\end{itemize}

\section*{Reproducibility}

Each step folder contains its preregistration, seal, OpenTimestamps proof, log, scripts, inputs,
readings and outputs. A seal is checked with \texttt{shasum -a 256} and its proof with the deposited
\texttt{verif\_ots.py}, which compares the proof with Bitcoin block headers. Random draws use fixed
seeds written in the preregistrations. The Retraction Watch file is not redistributed; its SHA-256 is
in the seal of the retraction step. Data, scripts, readings, full lists and proofs: \href{https://doi.org/10.5281/zenodo.23642239}{doi:10.5281/zenodo.23642239},
CC BY 4.0.

\section*{Competing interests}

The author is a co-founder of Loxyn, which develops the measurement instrument used here and
enterprise knowledge tools, has filed patent applications, and is preparing an application for the
Bpifrance Deep Tech label. The study was self-funded. The models tested were
chosen because they are available locally or from a French host.

\end{document}